\documentclass[amsmath,amssymb,prb,showpacs,twocolumn,superscriptaddress]{revtex4}

\usepackage{graphicx}
\newcommand{\w}{\omega}

\begin{document}
\makeatletter ¡¡¡¡
\newcommand{\rmnum}[1]{\romannumeral #1} ¡¡¡¡
\newcommand{\Rmnum}[1]{\expandafter\@slowromancap\romannumeral #1@}
\makeatother
\title{Nonequilibrium Green's function method for phonon-phonon interaction  and ballistic-diffusive thermal transport}
\author{Yong Xu}
\affiliation{Center for Advanced Study and Department of Physics,
Tsinghua University, Beijing 100084, China}
\affiliation{Department
of Physics and Centre for Computational Science and Engineering,
National University of Singapore, Singapore 117542,  Republic of
Singapore}
\author{Jian-Sheng Wang} \affiliation{Department of
Physics and Centre for Computational Science and Engineering,
National University of Singapore, Singapore 117542,  Republic of
Singapore}
\author{Wenhui Duan}
\email[\* Corresponding author. Email:\ ]{dwh@phys.tsinghua.edu.cn}
\affiliation{Center for Advanced Study and Department of Physics,
Tsinghua University, Beijing 100084, China}
\author{Bing-Lin Gu}
\affiliation{Center for Advanced Study and Department of Physics,
Tsinghua University, Beijing 100084, China}
\author{Baowen Li}
\affiliation{Department of Physics and Centre for Computational
Science and Engineering, National University of Singapore, Singapore
117542,  Republic of Singapore}
\affiliation{NUS Graduate School for
Integrative Sciences and Engineering, Singapore 117597, Republic of
Singapore}
\date{\today}

\begin{abstract}
Phonon-phonon interaction is systematically studied by
nonequilibrium Green's function (NEGF) formulism in momentum space
at finite temperatures. Within the quasi-particle approximation,
phonon frequency shift and lifetime are obtained from the retarded
self-energy. The lowest order NEGF provides the same phonon lifetime
as Fermi's golden rule. Thermal conductance is predicted by the
Landauer formula with a phenomenological transmission function. The
main advantage of our method is that it covers both ballistic and
diffusive limits and thermal conductance of different system sizes
can be easily obtained once the mode-dependent phonon mean free path
is calculated by NEGF. As an illustration, the method is applied to
two one-dimensional atom chain models (the FPU-$\beta$ model and the
$\phi^4$ model) with an additional harmonic on-site potential. The
obtained thermal conductance is compared with that from a
quasi-classical molecular dynamics method. The harmonic on-site
potential is shown to remove the divergence of thermal conductivity
in the FPU-$\beta$ model.
\end{abstract}

\maketitle

\section{\label{intro}Introduction}
More understanding of nanoscale thermal transport is required to
solve heat dissipation problem, which becomes important as the size
of electronic device decreases~\cite{cahill}. In recent years, many
experiments have been done to measure thermal conduction of
nanostructures~\cite{schwab,dli,shi,cnnt}. At the nanoscale, the
ballistic approximation is usually a good starting point for thermal
transport and a lot of theoretical research on ballistic thermal
transport have been reported~\cite{rego,yamamoto1,mingo-prl,ksaito}.
However, the ballistic approximation would lead to unphysical
results, such as infinite phonon mean free path, divergent thermal
conductivity and zero temperature gradient. In most cases we need to
go beyond ballistic limit to include effects of scattering for a
more realistic consideration. In fact, phonon-phonon interaction is
one of the significant factors for understanding and improving
thermal transport properties. Since the sizes of nanostructures are
comparable to phonon mean free path, thermal transport in these
systems is in the intermediate region between ballistic and
diffusive ranges. Nanoscale thermal transport has been studied
through many
approaches~\cite{jswang-epj,lepri,bli-jcp,bli-prb,jswang-prl,mingo-prb,yamamoto2,dhar,jswang-prb,mingo-negf,jswang-pre}.
Nevertheless, no satisfactory method has been available to deal with
thermal transport in the intermediate region.

Our work aims to provide an efficient way to study
ballistic-diffusive thermal transport including the effects of
phonon-phonon interaction. Here we propose a new formalism of
nonequilibrium Green's function (NEGF) to treat phonon-phonon
interaction  in momentum space at finite temperatures. It is known
that NEGF is usually used to study nonequilibrium problems, but its
formulism can be applied to equilibrium systems as well. While
Matsubara formalism~\cite{matsubara} is conventionally used for
equilibrium Green's function calculation at finite temperatures,
NEGF is an alternative that also works for such situations. In our
framework of NEGF, the analytical form of phonon self energy to any
order can be easily derived and phonon frequency shift and lifetime
are directly related to the retarded self energy. The phonon
frequency shift can also be accurately predicted by an effective
phonon theory based on the ergodic hypothesis (equipartition
theorem)~\cite{bli-epl2006,bli-epl2007,bli-pre2007}. For weak
phonon-phonon interaction, however, frequency shift is usually very
small and the major influence on thermal transport is from the
finite phonon lifetime. It will be shown that NEGF of lowest order
is equivalent to Fermi's golden rule when considering phonon
lifetime.

We also provide a new approach to study thermal transport using
NEGF. Our NEGF formalism gives phonon lifetime and mean free path.
Using a phenomenological transmission function determined from mean
free path in the Landauer formula, ballistic-diffusive thermal
transport can be studied~\cite{jswang-apl,jswang-epj,murphy}. Many
works have been done on thermal transport by
NEGF~\cite{yamamoto2,dhar,jswang-prb,mingo-negf,jswang-pre}. In
previous treatments, the system is situated at the center as a
junction with semi-infinite leads on the two sides. The central part
can not be too large due to the constraint of computational
capability. Here we compute phonon lifetime in a periodic system at
equilibrium, and feed the lifetime information to phonon
transmission function in the Landauer formula. Our approach is
computationally more efficient, but with a less rigorous treatment
of transmission function.

In Sec.~\Rmnum{2}, the general theory of NEGF is developed for
phonon-phonon interaction and ballistic-diffusive thermal transport.
As an application we employ the method to study two explicit models
in Sec.~\Rmnum{3}. A summary is made in Sec.~\Rmnum{4}.

\section{General theory}
In the harmonic approximation, a crystal can be described in terms
of non-interacting phonons. The concept of ``phonon'' remains valid
when the anharmonic contribution is small compared with the
harmonic. In this case, the quasi-particle approximation can be
made, and then the anharmonic effects give a complex shift to phonon
frequency~\cite{maradudin}: the real part shifts the value of
frequency while the imaginary part corresponds to phonon lifetime.

One of the earliest methods~\cite{klemens,srivastava,ygu} of
calculating phonon lifetime is based on Fermi's golden rule and the
Boltzmann-Peierls equation. However, those approaches can not be
systematically improved. In contrast, the Green's function method
provides a systematic way to consider phonon-phonon interaction at
finite temperatures. Phonon-phonon interaction has been studied
using equilibrium Green's function
method~\cite{maradudin,pathak,ipatova,monga,valle,procacci}. These
works can be roughly divided into two groups according to the
decoupling schemes used. For the first group,
 the temperature dependent
Green's function with Matsubara representation is employed based on
a general Wick's theorem which holds when the system is large
enough~\cite{maradudin,pathak,ipatova,monga}. A fictitious imaginary
time is used and an analytical continuation is needed to get the
retarded Green's function~\cite{AGD}. For the other group, a
decoupling scheme proposed by Valle and Procacci is
used~\cite{valle,procacci}. In detail, in the equation of motion for
the real time Green's function, thermal averages are replaced with
those appropriate for the harmonic Hamiltonian. The analytical
expression of phonon self energy including high order terms can be
easily evaluated by a computer-aided technique, despite of the lack
of a clear theoretical foundation.

We will employ the NEGF method as described in
Ref.~\onlinecite{jauho}. Its formalism has been generalized to treat
arbitrary nonequilibrium systems with arbitrary initial density
matrices by Wagner~\cite{wagner}. If interactions are adiabatically
switched on and an initial non-interacting Hamiltonian is used,
Wick's theorem is still valid for the contour ordered Green's
function~\cite{keldysh}. The adiabatic switch-on assumption leads to
a loss of information related to initial correlation, but is still
reasonable in most physical situations. We will use this assumption
and take the contour ordered Green's function to study phonon-phonon
interaction.


In this section, we develop the general theory of NEGF to deal with
phonon-phonon interaction and ballistic-diffusive thermal transport.
At first, a general Hamiltonian of anharmonic system is given, and
the contour ordered Green's function of phonon is defined. Then, two
schemes are provided to solve Green's function. One is to use the
equation of motion and recursive expansion rules. The other is to
apply Feynman rules for self energy and solve the Dyson equation.
Within the quasi-particle approximation, the retarded self energy
directly corresponds to phonon lifetime. Next we will compare the
NEGF method with Fermi's golden rule. Finally, we discuss how to
study
 ballistic-diffusive thermal transport with the information of
phonon mean free path.

\subsection{The Hamiltonian}
In a Taylor expansion of the potential energy $\Phi$ at
equilibrium configuration, terms higher than the second order
constitute anharmonic Hamiltonian:
\begin{equation}
H_A = \sum_{n=3,4, \cdots} {\frac{1}{n!}  \sum_{i_1,i_2,\cdots, i_n}
{\Phi_{i_1 i_2 \cdots i_n } u_{i_1} u_{i_2} \cdots u_{i_n}  }}.
\end{equation}
Herein, the system has $N$ unit cells and each cell contains $r$
atoms. We use $i \equiv (n_i,\tilde{i})$  $(n_i =1,2,\cdots, N)$
and $\tilde{i} \equiv (\kappa_i, \alpha_i)$ $(\kappa_i
=1,2,\cdots, r;\ \ \alpha_i=x,y,z)$ for convenience.
$u_i=\sqrt{M_{\kappa_i} }x_i$,
 $M_{\kappa_i}$ is the atom mass and $x_i$ is the displacement.
We expand $u_{i}$ as
\begin{equation}
u_i=\sum_{q} \frac{e^{i\mathbf{q}\centerdot \mathbf{R}_{n_i}}}
{\sqrt{N}} \epsilon_{\tilde{i}}^{j_q} (\mathbf{q}) A_q,   \ \ \
q\equiv (\mathbf{q},j_q) \label{u},
\end{equation}
where $j_q=1,2,\cdots,3r$ denotes the phonon branch, and
$\epsilon^{j_q}(\mathbf{q})$ is the normal mode eigenvector. The
phonon operator $A_q$ can be expressed by the usual phonon
annihilation operator $a_q$ and creation operator $a_q^{\dagger}$.
Introducing $\bar{q}\equiv (-\mathbf{q},j_q)$, we have
\begin{align}
A_q&=\sqrt{\frac{\hbar}{2\w_q} } (a_q + a_{\bar{q}}^{\dagger}), \ \ \
\ \ A_q^{\dagger}=A_{\bar{q}},
\end{align}
where $\w_q$ is the frequency of the normal mode $q$. Then the
harmonic Hamiltonian has the form
\begin{align}
H_0&=\sum_{q} \frac{1}{2} \dot{A}_q \dot{A}_{\bar{q}}  + \sum_{q}
\frac{1}{2}\w_q^2 A_q  A_{\bar{q}} , \\
\dot{A}_{q} &= -i \sqrt{ \frac{\hbar \w_q}{2} } (a_q
-a_{\bar{q}}^{\dagger}).
\end{align}
The anharmonic Hamiltonian becomes
\begin{align}
H_A=\sum_{n=3,4, \cdots} {\frac{1}{n} \sum_{q_1,q_2,\cdots, q_n}
{F_{q_1 q_2 \cdots  q_n } A_{q_1} A_{q_2} \cdots A_{q_n}  }},
\end{align}
where the \emph{n}-leg vertex is
\begin{align}
F_{q_1 q_2 \cdots q_n}=& \frac{\triangle (\mathbf{q}_1+\mathbf{q}_2+
\cdots +\mathbf{q}_n)} {(n-1)! N^{\frac{n}{2}-1}} \sum_{\tilde{i}_1}
\sum_{i_2,\cdots,i_n} \Phi_{i_1 i_2 \cdots i_n} \nonumber
\\
&\times  e^{i \mathbf{q}_2 \centerdot
(\mathbf{R}_{n_{i_2}}-\mathbf{R}_{n_{i_1}})} \cdots  e^{i
\mathbf{q}_n
\centerdot (\mathbf{R}_{n_{i_n}} -\mathbf{R}_{n_{i_1}}) } \nonumber \\
&\times \epsilon_{\tilde{i}_1}^{j_{q_1}}(\mathbf{q}_1)
\epsilon_{\tilde{i}_2}^{j_{q_2}}(\mathbf{q}_2)  \cdots
\epsilon_{\tilde{i}_n}^{j_{q_n}}(\mathbf{q}_n).
\end{align}
$\triangle(\mathbf{q})=1$ if $\mathbf{q}$ is zero or a reciprocal
lattice vector, otherwise $\triangle(\mathbf{q})=0$.

\subsection{Nonequilibrium Green's function}
We define the contour ordered Green's function of phonon as
\begin{equation}
G_{qq'}(\tau,\tau')=-\frac{i}{\hbar } \langle  \mathcal{T}_{\tau}
A_q(\tau) A_{q'}(\tau ')\rangle,
\end{equation}
where $\tau$ and $\tau'$ are defined in the complex plane, $ \langle
\cdots \rangle$ denotes the ordinary Gibbs statistical average, and
$\mathcal{T}_{\tau}$ is the contour-ordering operator. The contour
runs slightly above the real axis from $-\infty$ to $+\infty$, and
back to $-\infty$ slightly below the real axis. From our definition,
we have
\begin{equation}
G_{qq'}(\tau,\tau')=\delta_{q,\bar{q}'}G_{q \bar{q}}(\tau,\tau').
\label{Gaa}
\end{equation}
Note that only those Green's functions of the form $G_{q
\bar{q}}(\tau,\tau')$ are nonzero because of the orthogonality
between the normal modes.

The mapping of the contour ordered Green's function onto four
different normal Green's functions has been
discussed~\cite{jauho,jswang-pre,jswang-epj}. A label $\sigma=\pm 1$
is needed to distinguish whether $\tau$ is on the upper branch or on
the lower branch,
\begin{align}
G^{\sigma \sigma'}_{qq'}(t,t')=\lim_{\epsilon \rightarrow 0^+}
G_{qq'}(t+i \epsilon \sigma ,t'+i \epsilon \sigma').
\end{align}
$G^{++}=G^t$, $G^{--}=G^{\bar{t}}$,  $G^{+-}=G^<$ and
$G^{-+}=G^{>}$. There are another two types of Green's functions:
$G^r$ and $G^a$. The relations among these six Green's functions in
time and frequency domains have been presented in
Ref.~\onlinecite{jswang-pre}.

\subsection{The equation of motion}

Simple recursive expansion rules for the contour ordered Green's
function of phonon in coordinate space is originally proposed in
Ref.~\onlinecite{jswang-pre}. Differently, the phonon operators $
A_q$ used here are not Hermitian. However, we can show that the same
recursive expansion rules can be used by our phonon Green's function
defined in momentum space.

We define a general $n$-point Green's function
\begin{align}
&G_{q_1 q_2 \cdots q_n}(\tau_1,\tau_2,\cdots,\tau_n)
\nonumber\\
=&-\frac{i}{\hbar} \langle \mathcal{T}_{\tau} A_{q_1}(\tau_1)
A_{q_2}(\tau_2) \cdots A_{q_{n}}(\tau_{n})  \rangle \label{Gn}.
\end{align}
The phonon operators satisfy
\begin{align}
\ddot{A}_q &=-\frac{i}{\hbar}[\dot{A}_q,H] =-w_q^2 A_q \nonumber
\\&-\sum_{n=3,4,\cdots} { \sum_{q_2,q_3,\cdots, q_n}
{F_{\bar{q}q_2q_3\cdots q_n } A_{q_2} A_{q_3} \cdots A_{q_n} } }.
\label{A2}
\end{align}
From the definition,
\begin{align}
&F_{q_1 q_2 \cdots q_n} (\tau_1,\tau_2,\cdots,\tau_n) \nonumber \\
=&F_{q_1 q_2 \cdots q_n} \delta_{\sigma_1,\sigma_2} \delta(t_1 -
t_2)  \cdots \delta_{\sigma_1,\sigma_n} \delta(t_1 -
t_n)\sigma^{n-1}_1.
\end{align}
The equation for the contour ordered Green's function is
\begin{align}
& \frac{\partial^2}{\partial \tau^2}  G_{q q'}(\tau,\tau') = -
\delta_{q,\bar{q}'} \delta (\tau-\tau ') - w_q^2
G_{q,q'}(\tau,\tau')  \nonumber \\
&- \sum_{n=3,4,\cdots} \sum_{q_2,\cdots, q_n}  \int  \cdots
\int d \tau_2  \cdots d \tau_n \nonumber \\
& \times F_{\bar{q}q_2\cdots q_n }(\tau,\tau_2,\cdots,\tau_n)
G_{q_2\cdots q_n q'} (\tau_2,\cdots,\tau_n,\tau'). \label{G2}
\end{align}
The equation for the unperturbed Green's function is
\begin{align}
\frac{\partial^2}{\partial \tau'^2 }  G_{qq'}^0(\tau,\tau') =
-\delta_{q, \bar{q}'} \delta (\tau-\tau ') -w_{q'}^2 G_{q
q'}^0(\tau,\tau') \label{f2}.
\end{align}
Combining Eqs. (\ref{G2}) and(\ref{f2}), the two-point Green's
function can be expressed in terms of the free and higher order
ones as
\begin{align}
&G_{q q'}(\tau,\tau')=G_{q q'}^0(\tau,\tau')+ \sum_{n=3,4,\cdots}
\sum_{q_1,q_2,\cdots, q_n} \nonumber\\
&\Big[ \int \int \cdots \int d\tau_1 d\tau_2
 \cdots d\tau_n  \label{f3}G_{q q_1}^0(\tau,\tau_1) \nonumber\\
& F_{q_1 q_2 \cdots q_n } (\tau_1,\tau_2, \cdots, \tau_n) G_{q_2
\cdots q_n q'}(\tau_2,\cdots,\tau_n,\tau')\Big].
\end{align}
Repeating the above procedures, we can get the equation for higher
order Green's functions,
\begin{align}
&\ \ \ \ G_{q_1 q_2 \cdots q_n} (\tau_1 \tau_2 \cdots \tau_n)\nonumber\\
&= i  \hbar G^0_{q_1 q_2}(\tau_1,\tau_2) G_{q_3 q_4 \cdots q_n}
(\tau_3,\tau_4, \cdots,\tau_n) \nonumber\\
&+ i \hbar  G^0_{q_1 q_3}(\tau_1,\tau_3) G_{q_2 q_4 \cdots q_n}
(\tau_2,\tau_4 \cdots,\tau_n) +\cdots \nonumber\\
&+ i \hbar  G^0_{q_1 q_n}(\tau_1,\tau_n) G_{q_2 q_3 \cdots q_{n-1}}
(\tau_2,\tau_3 \cdots,\tau_{n-1}) \nonumber\\
 &+\sum_{m=3,4,\cdots} \sum_{q'_1,q'_2,\cdots, q'_m} \Big[ \int \int
\cdots \int d\tau'_1 d\tau'_2  \cdots d\tau'_m \nonumber\\
&\times G_{q_1 q'_1}^0(\tau_1,\tau'_1) F_{q'_1 q'_2 \cdots  q'_m }
(\tau'_1,\tau'_2,\cdots, \tau'_m) \nonumber\\
&\times G_{q'_2 \cdots q'_m q_2 \cdots
q_n}(\tau'_2,\cdots,\tau'_m,\tau_2,\cdots, \tau_n) \Big]. \label{f4}
\end{align}
Eqs. (\ref{f3}) and (\ref{f4}) are the equations of motion for the
contour ordered Green's function, which have the same form as
those in Ref.~\onlinecite{jswang-pre}. So the same recursive
expansion rules can be used here. One may conveniently implement
these rules in a computer program and expands the Green's function
to any order as one wish.

\subsection{Feynman rules for self energy}

The Dyson equation for our Green's function is
\begin{align}
&G_{q \bar{q}}(\tau,\tau')=G_{q\bar{q}}^0(\tau,\tau') \nonumber \\
&+\int \int d\tau_1 d\tau_2 G_{q \bar{q}}^0(\tau,\tau_1)
\Sigma_{\bar{q} q}(\tau_1,\tau_2) G_{q \bar{q}}(\tau_2,\tau').
\end{align}
For a steady or equilibrium system, it is more convenient to treat
problems in the frequency domain,
\begin{align}
G_{q \bar{q}}^{\sigma \sigma'}(\w)&= G_{q \bar{q}}^{0,\sigma
\sigma'}(\w)\nonumber \\
&+ \sum_{\sigma_1,\sigma_2} \sigma_1 \sigma_2 G_{q \bar{q}}^{0,
\sigma \sigma_1}(\w)  \Sigma_{\bar{q} q}^{\sigma_1 \sigma_2}(\w)
G_{q \bar{q}}^{\sigma_2 \sigma'}(\w).
\end{align}

\begin{figure}
\centering\includegraphics[width=0.45\textwidth]{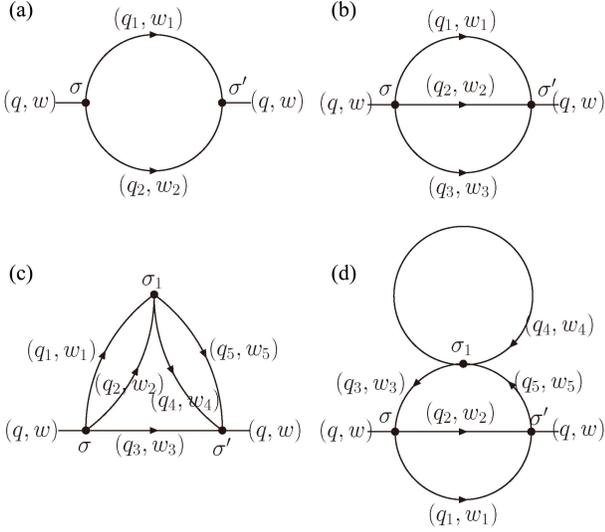}
\caption{Feynman diagrams for phonon self energy. }\label{fig1}
\end{figure}

Feynman diagrams for phonon self energy are the same as those in
Ref.~\onlinecite{jswang-pre}. Fig.~\ref{fig1} shows some lowest
order Feynman diagrams. In the following we summarize the Feynman
rules which are used to write the proper self energy
$\Sigma_{\bar{q} q}^{\sigma \sigma'}(\w)$.

(1) Draw all topologically distinct Feynman diagrams for the proper
self energy with two terminals on the left and right separately.
These diagrams should be connected and can not be separated into two
parts by cutting a single line.

(2) Draw an arrow on each line from left to right and label it with
new variable ($q_i,\w_i$). The line represents an unperturbed
Green's function $G_{q_i \bar{q}_i}^{0,\sigma_j \sigma_{j'}}(\w_i)$.
$\sigma_j,\sigma_{j'} = \pm 1$ are the variables of the vertices the
line connects, which will be discussed below. The variables of the
two terminals are both ($q,\w$). Using the relations between the six
Green's functions, $G_{q_i \bar{q}_i}^{0,\sigma_j
\sigma_{j'}}(\w_i)$ can be obtained from
\begin{equation}
G_{q_i \bar{q}_i}^{0,r} (\w_i) =\lim_{\delta \rightarrow
0^+}[(\w_i+i\delta)^2 - \w_{q_i}^2]^{-1}.
\end{equation}

(3) Label the vertices: each internal vertex with new variable
$\sigma_i$ ($\sigma_i = \pm 1$), the two terminal vertices with
$\sigma$ on the left and $\sigma'$ on the right. If only one
terminal vertex exists, the self energy is zero when $\sigma \neq
\sigma'$. Each $n$-leg vertex contributes a factor $F_{q_1/
\bar{q}_1, q_2/ \bar{q}_2, \cdots, q_n/ \bar{q}_n }$. Use
$\bar{q}_i$ in $F$ if the line ($q_i,\w_i$) goes into the vertex and
use $q_i$ in $F$ if the line goes out of the vertex. The momentum
conservation is automatically included in the factor.

 (4) At each vertex a factor of
$2\pi \delta \bigl( \sum_{i_{in}} {\w_{i_{in}}} -\sum_{i_{out}}
{\w_{i_{out}}} \bigr)$  is associated with energy conservation. The line
($q_{i_{in}}, \omega_{i_{in}}$) enters the vertex and the line
($q_{i_{out}}, \omega_{i_{out}}$) leaves the vertex. The two
terminal vertices only contribute one energy conservation factor.

(5) Multiply all the internal $\sigma$ variables and a coefficient
\begin{align}
c= &(-1)^{1+ \sum\limits_{n\geq3}N_n} (i)^{2+
\sum\limits_{n\geq3}\frac{n+2}{2}N_n } (\hbar)^{\sum\limits_{n\geq
3}\frac{n-2}{2}N_n } \nonumber \\
\times & \prod\limits_{n\geq 3}[(n-1)!]^{N_n} \frac{1}{S},
\end{align}
where $N_n$ is the number of $n$-leg vertices and $S$ is the
symmetry factor of the Feynman diagram.

(6) The sums or integrations are performed over all the internal
variables.

For example, the self energy of Fig.~\ref{fig1}(c) is
\begin{align}
&\Sigma_{\bar{q} q}^{\sigma \sigma'}(\w) = \sum_{q_1 \cdots q_5}
\sum_{\sigma_1} \int \frac{d\w_1}{2\pi}\cdots \int \frac{d\w_5}{2\pi}
(-54i \hbar^3) \sigma_1 \nonumber \\
&\times 2\pi \delta(\w-\w_1 -\w_2 -\w_3) 2\pi \delta(\w_1 +\w_2 -\w_4 -\w_5) \nonumber \\
&\times F_{\bar{q} q_1 q_2 q_3} F_{\bar{q}_1 \bar{q}_2 q_4 q_5}
F_{\bar{q}_3 \bar{q}_4 \bar{q}_5 q} G_{q_1 \bar{q}_1}^{0,\sigma
\sigma_1}(\w_1) G_{q_2 \bar{q}_2}^{0,\sigma \sigma_1}(\w_2) \nonumber \\
&\times G_{q_3 \bar{q}_3}^{0,\sigma \sigma'}(\w_3) G_{q_4
\bar{q}_4}^{0,\sigma_1 \sigma'}(\w_4)  G_{q_5 \bar{q}_5}^{0,\sigma_1
\sigma'}(\w_5). \label{self1}
\end{align}

In equilibrium system, only one of the six kinds of self energies is
independent. They have the same relations as the Green's functions.
For example, if $\Sigma_{\bar{q} q}^{++}(\w)=\Sigma_{\bar{q}
q}^{t}(\w)$ is calculated, the retarded self energy can be expressed
as
\begin{equation}
\Sigma_{\bar{q} q}^{r}(\w)= {\rm Re}[\Sigma_{\bar{q} q}^{t}(\w)] + i
\frac{ {\rm Im}[\Sigma_{\bar{q} q}^{t}(\w)]}{1+2f(\w)},
\end{equation}
where $f(\w)$ is the Bose-Einstein distribution.

\subsection{The retarded self energy and phonon lifetime}
In the quasi-particle approximation, the phonon frequency $\w_q$
suffers a complex shift $\Delta_q -i \Gamma_q$ due to phonon-phonon
interaction, where $\Gamma_q$ is the reciprocal phonon lifetime and
$2\Gamma_q$ is the full width at half-maximum (FWHM) of the phonon
peak in spectra. Then the frequency of the quasi-phonon is
$\tilde{\w}_q=\w_q+\Delta_q$.  The retarded phonon Green's function
can be written as
\begin{align}
G^r_{q \bar{q}}(\w)= \frac{1}{\w^2 - (\w_q + \Delta_q -i \Gamma_q)^2}.
\label{Gr2}
\end{align}
The Dyson equation for the retarded Green's function is
\begin{align}
G^r_{q \bar{q}}(\w)=\frac{1}{\w^2 - \w_q^2 - \Sigma_{\bar{q} q}^r(\w)}.
\label{Gr1}
\end{align}
The quasi-particle approximation is valid if
\begin{equation}
|\Delta_q - i \Gamma_q | \ll \w_q .\label{quasi}
\end{equation}
With this condition, from  Eqs.~(\ref{Gr2}) and (\ref{Gr1}) we get
the relations
\begin{align}
{\rm Re}\bigl[\Sigma_{\bar{q} q}^r(\w_q)\bigr] & \cong 2\w_q \Delta_q,
\\
{\rm Im}\bigl[\Sigma_{\bar{q} q}^r(\w_q)\bigr] & \cong -2\w_q \Gamma_q =-
\frac{2\w_q}{\tau_q}. \label{tau}
\end{align}
The imaginary part of the retarded self energy gives phonon
lifetime. The condition of the quasi-particle approximation becomes
$|\Sigma_{\bar{q} q}^r(\w_q)|\ll \w_q^2$ . Note that $\Sigma ^r
(-\w)=[\Sigma^r(\w)]^*$ implies ${\rm Re}[\Sigma ^r (-\w)]= {\rm
Re}[\Sigma^r(\w)]$ and ${\rm Im}[\Sigma ^r (-\w)]=-{\rm
Im}[\Sigma^r(\w)]$. The retarded self energy of ``$\w$-independent''
diagrams is real. To consider phonon lifetime, we only need to
calculate ``$\w$-dependent'' diagrams.

\subsection{The NEGF method and Fermi's golden rule}
Fermi's golden rule is widely used to calculate phonon lifetime in
previous studies. What is the relation and difference between the
NEGF method and Fermi's golden rule? This is the question to be
answered in this subsection.

Let's consider three-phonon interaction at first.  The lowest order
self energy contributed by the three-phonon interaction is described
by the Feymann diagram shown in Fig.~\ref{fig1}(a). Use the above
Feymann rules , we can write down the corresponding ``lesser'' self
energy as
\begin{align}
\Sigma_{\bar{q} q}^{<}(\w) = &\sum_{q_1 q_2} \int \frac{d\w_1}{2\pi}
\int \frac{d\w_2}{2\pi} (2i \hbar) 2\pi \delta(\w-\w_1 -\w_2)\nonumber \\
&\times F_{\bar{q} q_1 q_2} F_{q \bar{q}_1 \bar{q}_2} G_{q_1
\bar{q}_1}^{0,<}(\w_1) G_{q_2 \bar{q}_2}^{0,<}(\w_2).
\end{align}
The free ``lesser'' Green's function is
\begin{align}
G_{q \bar{q}}^{0,<}(\w)=\frac{-i \pi}{\w_q}[f(\w_q)\delta(\w -\w_q)
+ (f(\w_q)+1) \delta(\w+\w_q)].
\end{align}
The imaginary part of retarded self energy can be obtained through
the relation
\begin{align}
{\rm Im} [\Sigma_{\bar{q} q}^{<}(\w)]=2f(\w){\rm Im}
[\Sigma_{\bar{q} q}^{r}(\w)].
\end{align}
Using Eq.~(\ref{tau}), the reciprocal phonon lifetime can be
expressed as
\begin{align}
&\Gamma_q = \sum_{q_1 q_2} \frac{\pi \hbar}{4 \w_q \w_{q_1} \w_{q_2}
 } |F_{\bar{q} q_1 q_2}|^2  \{[f(\w_{q_1}) + f(\w_{q_2}) + 1] \nonumber \\
 & \times [\delta(\w - \w_{q_1} - \w_{q_2}) -\delta(\w + \w_{q_1} + \w_{q_2}) ] \nonumber \\
 &+ [f(\w_{q_1}) - f(\w_{q_2})][\delta(\w + \w_{q_1} - \w_{q_2}) -\delta(\w - \w_{q_1} + \w_{q_2}) ] \}.
\end{align}
On the other hand, we can use Fermi's golden rule to solve this
problem and obtain the same result of reciprocal phonon lifetime,
which is presented in Ref.~\onlinecite{Madelung}. Fermi's golden
rule is also equivalent to the lowest order NEGF for more than
three-phonon interaction. This conclusion can be proved in a similar
procedure as above and it will not be elaborated here.

In summary, NEGF of lowest order is equivalent to Fermi's golden
rule but the higher order terms of NEGF give additional corrections
to it. The NEGF method provides a more comprehensive and systematic
way than Fermi's golden rule to treat phonon-phonon interaction.

\subsection{Thermal transport from phonon mean free path}
Thermal conductance of a system at temperature $T$ can be given by
the Landauer formula as
\begin{align}
\sigma = \frac{1}{L}\sum_{q (v_{q}^x > 0)}\hbar \w_q v_{q}^x
\frac{\partial f(\w_q)}{\partial T}\, \Xi_q, \label{sigma}
\end{align}
where $L$ is the length of system in the direction of heat flow ($x$
direction here), $v_{q}^x$ is the phonon velocity in the $x$
direction, $f(\w_q)$ is the Bose-Einstein distribution, and $\Xi_q$
is the transmission function of phonon mode $q\equiv
(\mathbf{q},j_q)$. Eq.~(\ref{sigma}) is valid from one-dimensional
(1D) to three-dimensional (3D). Thermal conductivity has the form of
$\kappa=\sigma L/S$ in 3D systems, where $S$ is the cross section
area. And for quasi-1D systems, we define $\kappa=\sigma L$.

A phenomenological formula is proposed to calculate phonon
transmission function from the mean free
path~\cite{jswang-apl,jswang-epj,murphy}:
\begin{equation}
\Xi_q = (1+ L/l_q)^{-1} \label{Xi}
\end{equation}
with $l_q=v_q \tau_q$. When $L \ll l_q$, $\Xi_q=1$, corresponding to
the ballistic limit; when $L \gg l_q$, $\Xi_q= l_q/L$, thermal
conductivity is
\begin{align}
\kappa = \sum_{q (v_{q}^x > 0)}\hbar \w_q \frac{1}{V}\frac{\partial
f(\w_q)}{\partial T}\ v_{q}^x l_q,
\end{align}
where $V=L$ in 1D case and $V=L S$ for 3D case. It reproduces the
well-known Debye-Peierls formula for thermal transport. The formula
covers the range from ballistic regime to diffusive
regime~\cite{jswang-apl,jswang-epj}.

\section{Applications}
We apply the general theory developed to study explicit models in
this section. Firstly we introduce two 1D atom chain models and
investigate them by NEGF. Then a quasi-classical molecular dynamics
(QMD) method is applied to the same models and thermal conductances
from both methods are compared. Finally calculated results and
related discussion are provided.

\subsection{1D atom chain models}
The 1D atom chain models have been intensively studied. One of
remarkable problems is in what condition the thermal transport of a
1D system obeys Fourier law. The Fermi-Pasta-Ulam (FPU) $\beta$
model and the $\phi^4$ model are two example models which are
extensively used due to their simplicity. It is found that the
thermal conductivity in the FPU-beta model diverges with system size
due to the momentum conservation, while it is convergent in the
$\phi^4$ model~\cite{lepri-prl,aoki-prl,aoki-pla,bli-pre2000}.
Previous study indicates that the external potential plays a
determinant role for normal thermal
conduction~\cite{bli-pre1998,bli-pre2000}.

We will study  the FPU-$\beta$ model and the $\phi^4$ model with an
additional harmonic on-site potential. In the following, these two
models are denoted as model \Rmnum{1} and model \Rmnum{2},
respectively. The Hamiltonian of model \Rmnum{1} is
\begin{align}
H=\sum_{i} {\left[ \frac{ \dot{u}_i^2}{2} +
\frac{K}{2}(u_i-u_{i+1})^2 +\frac{K_0}{2}u_i^2 +
\frac{\beta}{4}(u_i-u_{i+1})^4 \right]} .
\end{align}
The Hamiltonian of model \Rmnum{2} has the form
\begin{align}
H=\sum_{i} {\left[ \frac{ \dot{u}_i^2}{2} +
\frac{K}{2}(u_i-u_{i+1})^2 +\frac{K_0}{2}u_i^2 + \frac{\mu}{4}u_i^4
\right]} .
\end{align}
They share the same harmonic Hamiltonian:
\begin{align}
H_0=\sum_{i} {\left[ \frac{ \dot{u}_i^2}{2} +
\frac{K}{2}(u_i-u_{i+1})^2 +\frac{K_0}{2}u_i^2 \right]} .
\end{align}
The dispersion relation of unperturbed phonon can be expressed as
$\w_q = \sqrt{K_0 + 2K(1-\cos q a)}$. We set the lattice constant $a
= 1$ \AA , $K = 1.0$ eV/(amu \AA$^2$) and $K_0 = 0.1 K$. These
parameters are chosen to be the same as those in
Ref.~\onlinecite{jswang-prl}, where model \Rmnum{2} is studied by
the QMD method. The group velocity is defined as $v_q=
\partial{\omega_q}/\partial{q}$. Since a harmonic
on-site potential is included in $H_0$, $v_{q = 0} = 0$ and
$\omega_q$ ranges from $0.31 \times 10^{14} \textrm{s}^{-1}$ to
$1.99 \times 10^{14} \textrm{s}^{-1}$.

For model \Rmnum{1}, we choose $\beta=1.0$ eV/(amu$^2$ \AA$^4$) and
the NEGF method provides converged result till $T$ = 600 K. For
model \Rmnum{2}, $\mu=1.0$ eV/(amu$^2$ \AA$^4$) is also tried , but
the NEGF method fails even for $T$ = 100 K. To get a converged
result, we select $\mu=0.05$ eV/(amu$^2$ \AA$^4$) and the NEGF
method would still work at the temperature higher than 1000 K. This
is partially because a small anharmonic parameter $\mu$ is chosen.
The calculations indicate that the NEGF method which is based on
perturbation expansion is only suitable for weak phonon-phonon
interactions.

The two models share the same set of Feynman diagrams for self
energy. Since their anharmonic parts are both quartic, only those
vertices with four legs will appear in the Feynman diagrams. The
4-leg vertex of model \Rmnum{1} has the form of
\begin{align}
&F_{q_1 q_2 q_3 q_4}=\Delta(q_1 +q_2 + q_3 +q_4)\frac{2 \beta }{N}[1- \cos{q_2 a} \\
\nonumber & - \cos{q_3 a} -\cos{q_4 a} + \cos{(q_2+q_3)a} + \cos{(q_2+q_4)a}\\
\nonumber& + \cos{(q_3+q_4)a}- \cos{(q_2+q_3+q_4)a}].
\end{align}
As $q_i$ ($i=1,2,3,4$) approaches zero,  $F_{q_1 q_2 q_3 q_4}$ goes
to 0. The 4-leg vertex of model \Rmnum{2} is
\begin{align}
F_{q_1q_2q_3q_4}=\Delta(q_1+q_2+q_3+q_4) \frac{\mu}{N}. \label{F}
\end{align}
Only ``$\omega$-dependent'' diagrams of lowest orders are considered
in our calculation. These include one diagram of O($\lambda^4$)
(Fig.~\ref{fig1}(b)) and two diagrams of O($\lambda^6$)
(Fig.~\ref{fig1}(c) and Fig.~\ref{fig1}(d)). Using the Feynman rules
developed above, the analytical form of self energy for these three
diagrams can be derived, and then be used in numerical calculation.
As described in the previous section, phonon lifetime and mean free
path can be obtained from the imaginary part of the retarded self
energy. Thermal transport properties can be given by the Landauer
formula.

\begin{figure}
\centering\includegraphics[width=0.45\textwidth]{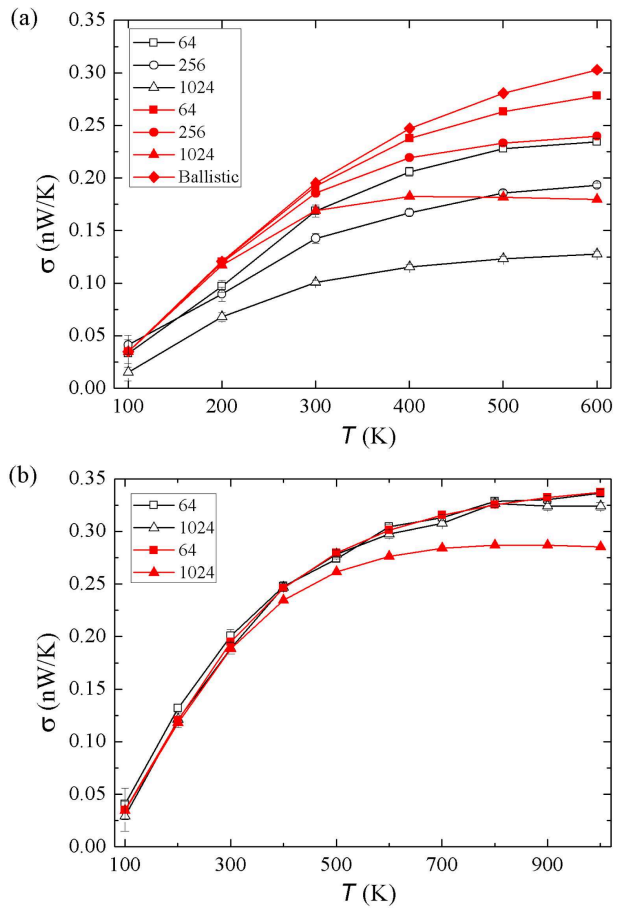}
\caption{(Color online) Comparison of thermal conductance ($\sigma$)
from different methods: NEGF (red solid symbols) and QMD (black open
symbols). (a) Thermal conductance of model \Rmnum{1} with the system
size $N = 64, 256, 1024$. Ballistic thermal conductance is also
provided by NEGF. (b) Thermal conductance of model \Rmnum{2} with
the system size $N = 64, 1024$.  }\label{fig2}
\end{figure}

\begin{figure*}
\centering\includegraphics[width=0.9\textwidth]{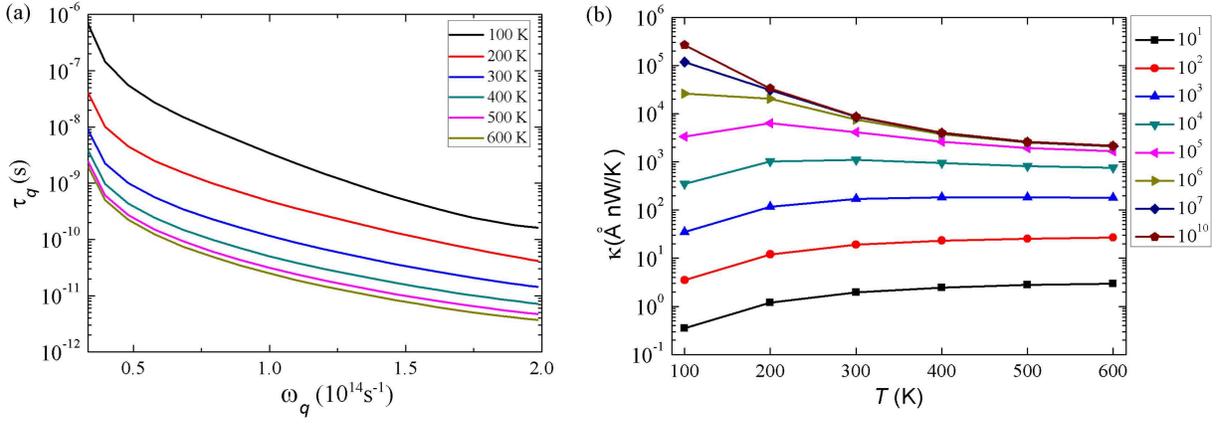}
\caption{(Color online) Results of Model \Rmnum{1}. (a) Phonon
lifetime $\tau_q$ at $T =$ 100 to 600 K. The nearly infinite
lifetimes of very long wavelength phonons are not presented in the
figure. (b) Thermal conductivity $\kappa$ with the system size $N=
10^1, 10^2, 10^3, 10^4, 10^5, 10^6, 10^7, 10^{10}$ at $T =$ 100 to
600 K. }\label{fig3}
\end{figure*}

\begin{figure*}
\centering\includegraphics[width=0.9\textwidth]{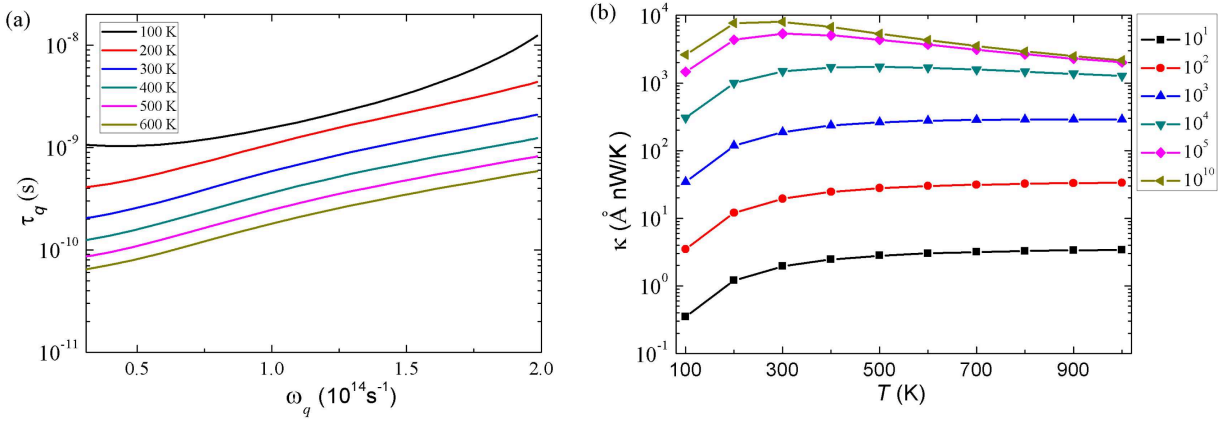}
\caption{(Color online) Results of Model \Rmnum{2}. (a) Phonon
lifetime $\tau_q$ at $T =$ 100 to 600 K. (b) Thermal conductivity
$\kappa$ with the system size $N= 10^1, 10^2, 10^3, 10^4, 10^5,
10^{10}$ at $T =$ 100 to 1000 K.}\label{fig4}
\end{figure*}

\subsection{Comparison of thermal conductance from NEGF and QMD}
Based on a generalized Langevin dynamics, the QMD method is
developed to study quantum thermal transport in
Ref.~\onlinecite{jswang-prl}, where the considered system consists
of a central junction part and two leads serving as heat baths. This
method uses quantum heat baths derived from Bose-Einstein statistics
and treats the central part classically. In
Ref.~\onlinecite{jswang-prl}, it is proved that the method can
produce correct results both in quantum ballistic and classical
diffusive limits.

The QMD method is employed to give thermal conductances of the two
models. $3 \times 10^8$ MD steps are used in the calculation with
the time step of $10^{-16}$s. The comparison of thermal conductance
with that from the NEGF method is presented in Fig.~\ref{fig2}  for
models \Rmnum{1} and \Rmnum{2}. Both methods give essentially the
same results at low temperatures. Deviations appear at high
temperatures. The NEGF method gives larger thermal conductance for
model \Rmnum{1} and smaller thermal conductance for model \Rmnum{2}.
It is not unexpected that they give quantitatively different
results. A quasi-classical approximation is made in the QMD method.
The NEGF method makes perturbation expansion and neglects higher
order terms which become important at high temperatures. Both
methods provide only approximate results, and it is still unclear
which method is superior. More work is needed to understand these
differences.

\subsection{Results and discussion}
The results of model \Rmnum{1} and model \Rmnum{2} are presented in
Fig.~\ref{fig3} and Fig.~\ref{fig4} respectively. The mode-dependent
phonon lifetime at different temperatures and the
temperature-dependent thermal conductivity with different system
sizes are shown .

The two models with different types of anharmonic potentials give
distinct properties of phonon lifetime. For model \Rmnum{1}, the
translational invariant quartic potential leads to zero retarded
self energy and infinite phonon lifetime for infinite long
wavelength mode, which is not shown in Fig.~\ref{fig3}(a). Long
wavelength phonon mode has long lifetime. In contrast, the long
wavelength phonon of model \Rmnum{2}, which experiences large
scattering due to the existence of quartic on-site potential, has
short lifetime.

Increasing temperature has two effects on thermal conductivity:
shorter phonon mean free path and more phonon excitation. The first
effect decreases and the second effect increases thermal
conductivity. When the system length is much smaller than the phonon
mean free path, the first effect has minor influence and thermal
conductivity would increase with increasing temperature. At high
temperatures, for a system whose length is comparable to the phonon
mean free path, the excitation of phonon modes becomes less
important to thermal transport and then thermal conductivity would
decrease with increasing temperature due to the shortening of phonon
mean free path. These are verified in Fig.~\ref{fig3}(b) and
Fig.~\ref{fig4}(b). For very large system at low temperatures, the
thermal conductivity of model \Rmnum{1} decreases as the temperature
increases, which is different from that of model \Rmnum{2}. This may
be explained by the fact that the temperature has larger influence
on the phonon mean free path in model \Rmnum{1} than in model
\Rmnum{2} at low temperatures.

\begin{figure}
\centering\includegraphics[width=0.45\textwidth]{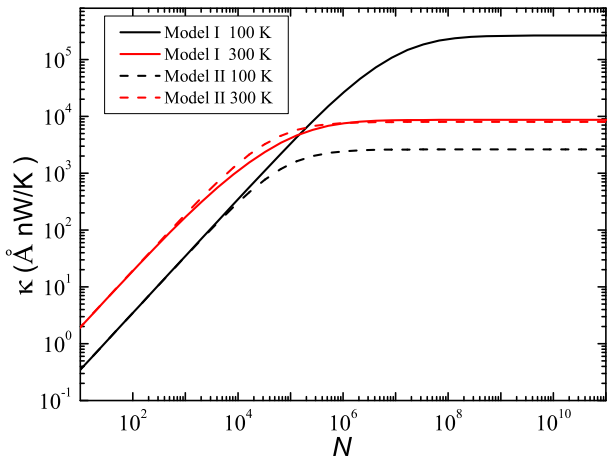}
\caption{(Color online) Thermal conductivity $\kappa$ of different
system sizes for model \Rmnum{1} and model \Rmnum{2} at $T =$ 100,
300 K.}\label{fig5}
\end{figure}

One of the advantages of our method is that once the mode-dependent
phonon mean free path is obtained, thermal conductivity of different
system sizes can be obtained easily. It would be interesting to show
the size-dependent behavior of thermal conductivity, which is
presented in Fig.~\ref{fig5}. When the system length is much smaller
than the phonon mean free path (\emph{i.e.}, the ballistic regime)
thermal conductance does not change with system size, and thermal
conductivity increases linearly with the system size. When the
system length increases to much larger than the phonon mean free
path (of order of $10^6$ \AA\  in model \Rmnum{2} at 100 K)
(\emph{i.e.}, the diffusive regime), thermal conductivity will cease
to change with system size. A ballistic-diffusive transition with
increasing system size is shown in Fig.~\ref{fig5} clearly.

The FPU-$\beta$ model, model \Rmnum{1} with  $K_0 = 0$, has been
widely known to have divergent thermal conductivity. With an
additional harmonic on-site potential, we get a finite bulk-limit
thermal conductivity for model \Rmnum{1} as shown in
Fig.~\ref{fig5}. How does a nonzero $K_0$, the harmonic on-site
potential, induce different behavior of thermal conductivity? An
analysis on long wavelength phonon modes will answer this question.
As $q$ approaches zero, $F_{q q_i q_j q_k} \propto q$ and
$\Sigma_{\bar{q} q}^r (\omega_q) \propto q^2$. When $K_0 = 0$,
$\omega_{q} \propto q$, $v_q \rightarrow$ constant, $\tau_q \propto
q^{-1}$ and $l_q \propto q^{-1}$. Thermal conductivity contributed
by long wavelength phonon modes would diverge with increasing system
size.  When $K_0 \neq 0$, $\omega_{q} \rightarrow$ constant, $v_q
\propto q$, $\tau_q \propto q^{-2}$ and $l_q \propto q^{-1}$. The
quadratic on-site potential breaks translational invariance and the
phonon modes with very long wavelength have nearly zero group
velocity. Though those phonon modes has nearly infinite mean free
path, it can be shown analytically that the thermal conductivity
contributed by them does not diverge with increasing system size for
$K_0 \neq 0$. Other phonon modes which have finite mean free paths
also contribute a finite thermal conductivity. So finite thermal
conductivity is obtained in bulk limit for model \Rmnum{1}.

Comparing the results of the two models may give some information on
how an anharmonic on-site potential influences thermal transport. A
quartic on-site potential is included in model \Rmnum{2}. The
anharmonic parameter $\mu$ of model \Rmnum{2} is 20 times as small
as that of model \Rmnum{1}. However, model \Rmnum{2} gives much
shorter phonon lifetime for low frequency modes, which are very
important for thermal transport especially at low temperatures. As
shown in Fig.~\ref{fig5}, the bulk-limit thermal conductivity of
model \Rmnum{2} is much smaller than that of model \Rmnum{1} at 100
K. The comparison indicates that even a small quartic on-site
potential can largely decrease the phonon lifetimes of long
wavelength modes and significantly affect thermal transport
especially at low temperatures. Our results confirm that the
external potential plays a determinant role in thermal transport.

\section{Summary}
We have provided a new approach to study phonon-phonon interaction
and ballistic-diffusive thermal transport by the NEGF method and the
Landauer formula. A new formalism of NEGF has been developed to
systematically study phonon-phonon interaction in momentum space at
finite temperatures in equilibrium. Using a phenomenological
transmission function, which can be obtained from the mode-dependent
phonon mean free path given by the NEGF formalism, the Landauer
formula predicts thermal transport properties from ballistic region
to diffusive region. Our approach is efficient for investigating
ballistic-diffusive thermal transport in weak interaction
situations, where little additional computational effort is needed
when the system size changes. As an application, we have
investigated two 1D atom chain models. The results obtained are
qualitatively agree with those by the QMD method. It is found that
an additional harmonic on-site potential in the FPU-$\beta$ model
could remove the divergence of thermal conductivity and a small
quartic on-site potential can largely reduce the phonon lifetimes of
long wavelength modes. The results confirm that the external
potential plays an important role in thermal
transport~\cite{bli-pre1998,bli-pre2000}.

\begin{acknowledgments}
The authors are grateful to Gang Zhang, Jingtao L\"u, Lifa Zhang and
Xiang Wu for discussions. This work is supported by an endowment
fund grant R-144-000-222-646 from NUS,  the National Natural Science
Foundation of China (Grant. 10547002) and the Ministry of Science
and Technology of China (Grant Nos. 2006CB605105 and 2006CB0L0601).
J.-S.  W.  is  supported in  part  by  an  NUS  Faculty  Research
Grant  R-144-000-173-101/112.

\end{acknowledgments}


\begin{thebibliography}{99}
\bibitem{cahill} D. G. Cahill, W. K. Ford, K. E. Goodson, G. D. Mahan, A.
Majumdar, H. J. Maris, R. Merlin, and S. R. Phillpot, J. Appl. Phys.
\textbf{93}, 793 (2003).

\bibitem{schwab} K. Schwab, E. A. Henriksen, J. M. Worlock, and M. L. Roukes,
Nature (London) \textbf{404}, 974 (2000).

\bibitem{dli} D. Li, Y. Wu, R. Fan, P. Yang, and A. Majumdar, Appl. Phys. Lett.
\textbf{83}, 3186 (2003).

\bibitem{shi} L. Shi, D. Li, C. Yu, W. Jang, D. Kim, Z. Yao, P. Kim, and A. Majumdar, J. Heat Transfer \textbf{125}, 881 (2003).

\bibitem{cnnt} J. Hone, M. Whitney, C. Piskoti, and
A. Zettl, Phys. Rev. B ~\textbf{59}, R2514 (1999); P. Kim, L. Shi,
A. Majumdar, and P. L. McEuen, Phys. Rev. Lett. ~\textbf{87}, 215502
(2001); M. Fujii, X. Zhang, H. Xie, H. Ago, K. Takahashi, T. Ikuta,
H. Abe, and T. Shimizu, Phys. Rev. Lett. ~\textbf{95}, 065502
(2005); E. Pop, D. Mann, Q. Wang, K. E. Goodson, and H. Dai, Nano
Lett. \textbf{6}, 96 (2006).

\bibitem{rego} L.G.C. Rego and G. Kirczenow, Phys. Rev. Lett. \textbf{81}, 232
(1998).

\bibitem{yamamoto1} T. Yamamoto, S. Watanabe, and K. Watanabe, Phys. Rev.
Lett. \textbf{92}, 075502 (2004).

\bibitem{mingo-prl} N. Mingo and D. A. Broido, Phys. Rev. Lett. \textbf{95}, 096105
(2005).

\bibitem{ksaito} K. Saito, J. Nakamura, and A. Natori, Phys. Rev. B
\textbf{76}, 115409 (2007).


\bibitem{jswang-epj} J.-S. Wang, J.Wang and J. T. L\"u, Eur. Phys. J. B \textbf{62}, 381 (2008).

\bibitem{lepri} S. Lepri, R. Livi, and A. Politi, Phys. Rep. \textbf{377}, 1 (2003).

\bibitem{bli-jcp} G. Zhang and B. Li, J. Chem. Phys. \textbf{123}, 114714 (2005).

\bibitem{bli-prb} L. H. Liang and B. Li, Phys. Rev. B \textbf{73}, 153303 (2006).

\bibitem{jswang-prl} J.-S. Wang, Phys. Rev. Lett. \textbf{99}, 160601 (2007).

\bibitem{mingo-prb} N. Mingo and L. Yang, Phys. Rev. B \textbf{68}, 245406 (2003).

\bibitem{yamamoto2} T. Yamamoto and K. Watanabe, Phys. Rev. Lett.
\textbf{96}, 255503 (2006).

\bibitem{dhar} A. Dhar and D. Sen, Phys. Rev. B \textbf{73}, 085119 (2006).

\bibitem{jswang-prb} J.-S. Wang, J. Wang, and N. Zeng,
Phys. Rev. B \textbf{74}, 033408 (2006).

\bibitem{mingo-negf} N. Mingo, Phys. Rev. B \textbf{74}, 125402 (2006).

\bibitem{jswang-pre} J.-S. Wang, N. Zeng, J. Wang, C.K. Gan, Phys. Rev. E \textbf{75},
061128 (2007).

\bibitem{matsubara}T.~Matsubara, Progr. Theor. Phys. \textbf{14}, 351
(1955).

\bibitem{bli-epl2006} N. Li, P. Tong, and B. Li, Europhys. Lett. \textbf{75}, 49 (2006).

\bibitem{bli-epl2007} N. Li and B. Li, Europhys. Lett. \textbf{78}, 34001 (2007).

\bibitem{bli-pre2007} N. Li and B. Li, Phys. Rev. E \textbf{76}, 011108 (2007).

\bibitem{jswang-apl} J. Wang and J.-S. Wang, Appl. Phys. Lett. \textbf{88}, 111909
(2006).


\bibitem{murphy} P. G. Murphy and J. E. Moore, Phys. Rev. B \textbf{76}, 155313
(2007).

\bibitem{maradudin} A. A. Maradudin and A. E. Fein, Phys. Rev. \textbf{128}, 2589
(1962).

\bibitem{klemens} M. Roufosse and P. G. Klemens, Phys. Rev. B \textbf{7}, 5379 (1973).

\bibitem{srivastava} S. P. Hepplestone and G. P. Srivastava,
Phys. Rev. B \textbf{74}, 165420 (2006).

\bibitem{ygu} Y. Gu and Y. Chen, Phys. Rev. B \textbf{76}, 134110 (2007).

\bibitem{pathak} K. N. Pathak, Phys. Rev. \textbf{139}, A1569
(1965).

\bibitem{ipatova} I. P. Ipatova, A. A. Maradudin, and R. F. Wallis, Phys. Rev. \textbf{155}, 882
(1967).

\bibitem{monga} M. R. Monga and K. N. Pathak, Phys. Rev. B \textbf{18}, 5859
(1978).

\bibitem{valle} R.~G. DellaValle, P.~Procacci, Phys.~Rev.~B \textbf{46}, 6141 (1992).

\bibitem{procacci}P. Procacci, G. F. Signorini, and R.~G. DellaValle, Phys. Rev. B \textbf{47}, 11124
(1993).

\bibitem{AGD} A. A. Abrikosov, L. P. Gorkov, and I. E. Dzyaloshinski, \textsl{Quantum Field Theoretical Methods in Statistical Physics} (Pergamon Press,
1965).

\bibitem{jauho} H. Haug and A. P. Jauho, \textsl{Quantum Kinetics in
Transport and Optics of Semiconductors} (Springer, Berlin, 1996).

\bibitem{wagner} M. Wagner, Phys. Rev. B \textbf{44}, 6104 (1991).

\bibitem{keldysh} L. V. Keldysh, Sov. Phys. JETP. \textbf{20}, 1018 (1965).

\bibitem{Madelung} O. Madelung, \textsl{Introduction to Solid-State Theory} (Springer-Verlag Berlin Heidelberg, 1978).

\bibitem{lepri-prl} S. Lepri, R. Livi, and A. Politi, Phys. Rev. Lett. \textbf{78}, 1896
(1997).

\bibitem{bli-pre2000} B. Hu, B. Li, and H. Zhao, Phys. Rev. E ~\textbf{61}, 3828
(2000).


\bibitem{aoki-pla} K. Aoki and D. Kusnezov, Phys. Lett. A \textbf{265}, 250
(2000).

\bibitem{aoki-prl} K. Aoki and D. Kusnezov, Phys. Rev. Lett. \textbf{86}, 4029 (2001).


\bibitem{bli-pre1998} B. Hu, B. Li, and H. Zhao, Phys. Rev. E
~\textbf{57}, 2992 (1998).




\end{thebibliography}
\end{document}